# Creation of high mobility two-dimensional electron gases via strain induced polarization at an otherwise nonpolar complex oxide interface

*Yunzhong Chen,\*[1] Felix Trier,[1] Takeshi Kasama,[2] Dennis V. Christensen,[1] Nicolas Bovet,[3] Zoltan I. Balogh,[2] Han Li[1], Karl Tor Sune Thydén[1], Wei Zhang[1], Sadegh Yazdi,[2] Poul Norby[1], Nini Pryds,[1] and Søren Linderoth[1]*

[1]Department of Energy Conversion and Storage, Technical University of Denmark, Risø campus, 4000 Roskilde, Denmark

[2]Center for Electron Nanoscopy, Technical University of Denmark, 2800 Lyngby, Denmark

[3]Nano-Science Center, Department of Chemistry, University of Copenhagen, 2100 Copenhagen, Denmark

**ABSTRACT**: The discovery of two-dimensional electron gases (2DEGs) in $SrTiO_3$-based heterostructures provides new opportunities for nanoelectronics. Herein, we create a new type of oxide 2DEG by the epitaxial-strain-induced polarization at an otherwise nonpolar perovskite-type interface of $CaZrO_3/SrTiO_3$. Remarkably, this heterointerface is atomically sharp, and exhibits a high electron mobility exceeding 60,000 $cm^2V^{-1}s^{-1}$ at low temperatures. The 2DEG carrier density exhibits a critical dependence on the film thickness, in good agreement with the polarization induced 2DEG scheme.

---

\* Corresponding Author. Email:yunc@dtu.dk. Phone: +45 4677 5614.

**KEYWORDS**: Complex oxide interfaces, oxide electronics, two-dimensional electron gases, strain induced polarization.

Atomically engineered complex oxide heterostructures exhibit a variety of exotic interfacial properties because of strong interactions among the spin, charge, and orbital freedoms as well as lattice vibrations. One particular example is the emergence of high mobility two-dimensional electron gases (2DEGs) at the interface between two oxide insulators,[1,2] one of which is SrTiO$_3$ (STO), the basis material of oxide electronics. These complex oxide 2DEGs consist of strongly coupled electrons and give rise to a rich set of physical phenomena[3-5], for example, superconductivity[6,7], magnetism[8,9], and tunable metal-insulator transitions on nanoscale,[10,11] providing new opportunities for nanoelectronics and mesoscopic physics.[3-5] To date, the creation of complex oxide 2DEGs can mainly be divided into three groups: (1) electronic reconstruction resulting from the polar discontinuity at a polar-nonpolar oxide interface. Such interfacial polar discontinuity requires the presence of a polar oxide, that consists of alternating positively and negatively charged sublayers, epitaxially grown on nonpolar SrTiO$_3$, such as LaAlO$_3$/SrTiO$_3$ (LAO/STO)[1]; (2) Creation of oxygen vacancies at the bare STO surface or in STO-based heterostructures having interfacial redox reactions[12-15], such as vacuum cleaved STO[12-14] or the system of amorphous-LaAlO$_3$/SrTiO$_3$ ($a$-LAO/STO)[15]; (3) 2DEGs by delta doping of STO, typically sandwiching a Nb or La-doped STO thin layer between two non-doped STO layers.[16] Among these, the polarity issue, particularly in the LAO/STO system, has drawn the most attention due to the possibility of an intrinsic doping of STO by electronic reconstruction without the typical disorder caused by chemical doping.[17] However, to compensate the interfacial polar discontinuity, competing mechanisms have often been proposed to occur and obscure the sought-after electronic reconstruction.[18-20] For example, the formation of oxygen vacancies and/or the occurrence of cation intermixing have been shown to play important roles in these STO-based 2DEGs systems[18,19]. To achieve full understanding of complex oxide 2DEGs and to further improve their physical properties such as carrier mobilities, it is necessary to explore perovskite interfaces where competing mechanisms can be isolated distinctly. In this vein, it is nontrivial to investigate the isopolar or nonpolar perovskite heterostructures. Compared to polar oxide interfaces, the isopolar or nonpolar interfaces can not only avoid the intrinsic polar discontinuity issue at the interface[21] but can also avoid the possible cation intermixing induced charge compensation mechanisms through formation of donor-acceptor antisite defect pairs.[20]

Besides interfacial polar discontinuity, spontaneous and/or piezoelectric polarizations have been found to result in high-mobility 2DEGs in both traditional semiconductor materials such as III-V semiconductor AlGaN/GaN compound heterostructures[22] and the binary oxide heterostructures of ZnMgO/ZnO[23,24]. In these systems, 2DEGs carriers normally originate from the ionized surface donor state. The spontaneous and/or piezoelectric polarization in the top film provides the driving force to collect the electrons in the quantum well.[22-24] Though the spontaneous and/or piezoelectric polarizations of ferroelectric oxides have been used to control the physical properties of 2DEGs at the LAO/STO interface [25, 26], their role as the origin for formation of complex oxide 2DEGs has not been reported previously, to the best of our knowledge. Herein, we report for the first time a high mobility 2DEG at an otherwise nonpolar perovskite-type interface of (001) $CaZrO_3$/STO (CZO/STO) via strain induced polarization, as demonstrated in Fig. 1a. In contrast to the polar nature of LAO films[1], the formal valence states of CZO can be assigned as $Ca^{2+}$, $Zr^{4+}$ and $O^{2-}$, in the simple ionic limit. Therefore, the (001) CZO consists of alternating charge neutral stacks of $(Ca^{2+}O^{2-})^0$ and $(Zr^{4+}O^{2-}_2)^0$, similar to the case of nonpolar (001) STO. Nevertheless, we find that compressive strain can induce an electric polarization in the CZO epitaxial thin films, consistent with previous reports.[27, 28] Remarkably, such strain induced polarization could well account for the 2DEGs at our CZO/STO heterointerface, as demonstrated by a critical dependence of the interfacial conduction on the CZO film thickness. Moreover, the nominally-nonpolar CZO/STO interface gives rise to an extremely high electron mobility exceeding 60000 $cm^2V^{-1}s^{-1}$ at 2 K, which is among one of the highest mobilities achieved for perovskite-type interfaces.[2,29,30] These findings provide new opportunities for oxide electronics.

The CZO (pseudocubic $a$=4.012 Å)[31] exhibits a good lattice match with STO ($a$=3.905 Å). Their lattice mismatch is +2.67%, comparable to the mismatch of -2.79% in the LAO/STO system, where the positive and negative values indicate a compressive and tensile strain in the epitaxial film, respectively. Under optimized conditions, the CZO films deposited by pulsed laser deposition (PLD) can be epitaxially grown on the (001) $TiO_2$-terminated STO substrates within a layer-by-layer two-dimensional growth mode, as confirmed by the presence of periodic intensity oscillations of the reflection high-energy electron diffraction (RHEED) pattern monitored *in-situ* during film growth (Supporting information, Fig.S1). Both RHEED intensity oscillations and sharp RHEED patterns can persist up to a film thickness over 50 unit cells (uc), suggesting high quality film growth. A terrace surface of the grown heterostructure is detected by atomic force microscopy (AFM), which shows a regular step height of 0.4 nm (Fig. 1b). High-resolution X-ray diffraction (XRD) further confirms the epitaxial growth of the (001) CZO film on the (001)-oriented STO, without the presence of any impurity phases. The XRD reciprocal space maps (RSM) measurements in the vicinity of the (002) and (-103) perovskite Bragg peaks, as demonstrated in Figs. 1c and d, respectively, give direct evidence that the films are under compressive biaxial strain, as expected.

We further investigated the atomic structure and interface chemistry of our CZO/STO heterostructures by an aberration corrected scanning transmission electron microscopy (STEM) in combination with electron energy-loss spectroscopy (EELS). Figure 2a shows a high-angle annular dark field (HAADF) STEM image of a CZO/STO sample with the CZO layer of approximately 50 uc (~20 nm). The CZO film is found to be coherent with the STO substrate with no obvious defects or dislocations at the interface. The averaged line profiles (Fig. 2b) across the interface indicate that most of the cation interdiffusion is confined to within 1-2 unit cell on either side of the interface, much sharper than the spinel/perovskite interface of GAO/STO grown under similar conditions.[2] The CZO/STO interface is therefore among one of the sharpest hitherto-investigated perovskite-type oxide interfaces.[18,32,33] Further STEM investigations indicate that the epitaxial CZO film is strained compressively on STO with $a≈b≈3.905$ Å and $c≈4.124$ Å, consistent with the XRD data shown in Fig.1. In a similar zirconate system of $SrZrO_3/SrTiO_3$, the compressive strain has been reported to result in ferroelectricity in its superlattices.[27] Theoretical calculations by first principle density functional theory indicate that such strain induced polarization originates from a lattice distortion, which is characterized by a prediction that all the cations move towards the interface with the $Zr^{4+}$ exhibiting the largest relative displacement.[28] Therefore, we also investigated the lattice distortion of our CZO films in the vicinity of the interface. Remarkably, compared to the $CaZrO_3$ film that is more than 10 uc away from the interface, we observed unambiguously that both the $Ca^{2+}$ and $Zr^{4+}$ cations move toward the interface in the first 7 uc CZO layers. As shown in Fig. 2c, the $Zr^{4+}$ cation in the first $ZrO_2$ sublayer exhibits the largest relative movement of 0.035 nm to the interface, while such trend degrades quickly in the very beginning 3 unit cells of CZO films. In contrast, the displacement of the $Ca^{2+}$ cations persists up to the first 7 uc layers. The above lattice displacement implies a polarization pointing towards the interface, given that there is a negligible lattice distortion for the oxygen sublattice as illustrated in Fig. 2d.

The stoichiometric CZO and STO are both band gap insulators, with the energy band gap of 4.1eV and 3.2 eV, respectively. Nevertheless, we measured highly mobile metallic conduction at the epitaxial CZO/STO heterointerface. Figure 3a shows the typical temperature dependent sheet resistance, $R_s$, for our CZO/STO heterostructures at different film thickness, $t$. All of our CZO/STO heterostructures show good metallic behavior as long as the CZO film thickness $t$ is higher than 6 uc. Note that the samples are highly insulating at $t \leq 6$ uc. This rules out the possibility of thermal reduction of STO as a contribution to the interface conduction.[34] This critical thickness $t_c$=6 uc (~2.4 nm) for the occurrence of interface conduction is slightly higher than that of the LAO/STO interface ($t_c$~1.6 nm)[35]. However, in distinct contrast to the LAO/STO system, where the carrier density normally exhibits a carrier freezing out behavior at $T$=100 K, the CZO/STO system exhibits an almost constant carrier density as a function of temperature (Fig.3b), similar to the high mobility spinel/perovskite GAO/STO interface[2, 36]. Moreover, the CZO/STO system distinguishes from the other complex oxide 2DEGs by two characteristic properties: Firstly, we observe two jumps in the carrier density (Fig. 3b and also Fig.4a) upon increasing film thickness. Besides the occurrence of interface conduction at around $t_c$ =6 uc for the insulator-metal transition, we observed another unexpected conduction jump in the carrier density by more than 40 times at $t$ =15 uc, from approximately $n_s$=0.4-2.0 ×$10^{13}$ cm$^{-2}$ at $t \leq 14$ uc to $n_s$=0.4-1.1 ×$10^{15}$ cm$^{-2}$ at $t \geq 15$ uc. Secondly, despite the strong dependence of $n_s$ on $t$, most of the CZO samples show very high electron mobilities much larger than 20 000 $cm^2V^{-1}s^{-1}$ at 2 K (the bulk mobility of STO)[37], which is nearly regardless of the CZO film thickness (Fig. 3c). For example, we obtained a $\mu = 5.3 \times 10^4$ $cm^2V^{-1}s^{-1}$ and $\mu = 6.1 \times 10^4$ $cm^2V^{-1}s^{-1}$ at 2 K for samples with $t=10$ uc and $t=18$ uc, respectively, which represents the record mobility for perovskite-type interfaces and are also among the highest mobilities for complex oxide interfaces (supporting information, Table S1)[2,29,30].

The metallic conduction at the nominally-nonpolar CZO/STO heterointerface is remarkable, since two of the most discussed conduction mechanism for the polar LAO/STO interface, intrinsic polar discontinuity and La-doping induced $n$-type conduction, are both ruled out here. On the other hand, besides the presence of strain-induced polarization in the CZO epitaxial thin films, the CZO/STO heterointerface meets also the thermodynamic criterion for interfacial redox reaction: the $B$ site metal (Zr for CZO system) locates in the region of the formation heat of metal oxide, $\Delta H_f^O < -350$ kJ/(mol O) and the work function of the metals, 3.75 eV $< \varphi <$ 4.4 eV.[38] The redox reaction at the interface of STO-based heterostructures can result in oxygen vacancies dominated metallic conduction.[15] To explore the role of oxygen vacancies in the CZO/STO heterostructures, we performed angle-resolved X-ray photoelectron spectroscopy (XPS) measurements. Generally, the presence of oxygen vacancies in these STO-based oxide 2DEG heterostructures is indicated by the fact that the content of $3d$ electrons, $Ti^{3+}$ on the STO side, deduced by XPS measurements is much larger than the carrier density obtained from transport measurements.[2,15] Remarkably, in dramatic difference with both crystalline LAO/STO[39] and GAO/STO systems[2], negligible $Ti^{3+}$ signal in the $2p_{3/2}$ core-level spectra is detected in our CZO/STO heterostructures at $t \leq 14$ uc (Supporting information, Fig. S2). This strongly suggests that the content of oxygen vacancies in our CZO/STO samples is negligible for $t \leq 14$ uc (approximately the detection limit of XPS). Such conclusion is further supported by the fact that all our samples can survive the post annealing in 1 bar pure $O_2$ at 100 ºC for over 5 hours (Supporting information, Fig. S3), with no obvious change in the interface conduction. Therefore, the 2DEG at the CZO/STO ($t \leq 14$ uc) interface results predominantly from the strain induced polarization, $P_{CZO}$, in the CZO films. Note that although the content of the oxygen vacancies in the CZO/STO ($t \leq 14$ uc) are strongly suppressed, the 2DEG carriers originating from the polarization-induced electronic reconstruction still locate preferably on the STO side in the region proximate to the interface. This is confirmed by our observation that the $e_g/t_{2g}$ ratio of the $L_{2,3}$ edge in the EELS spectra of STO is decreased close to the interface with respect to the ones further in the bulk STO (Supporting information, Fig. S4). The polarization in the epitaxial CZO films could result from the relative displacements of the cations and anions under compressive strain, as shown in Fig. 2d. If we assume a constant polarization, $P_{CZO}$, in the CZO film, this polarization will lead to the presence of a macroscopic electric field $E_{CZO} = P_{CZO}/\varepsilon_0\varepsilon_{CZO}$ across CZO films ($\varepsilon_{CZO}$ is the relative permittivity of CZO), which will bend the electronic bands. As illustrated in Fig. 4b, for ultrathin CZO films with a thickness below $t_c$, the ionized donor surface states, which are sufficiently deep and below the conduction band of CZO with an energy $E_D$, lie far below the conduction band of STO. In this case, no 2DEG is formed. Increasing the CZO film thickness normally lifts the surface donor states. At the critical thickness, the surface donor energy reaches the bottom of the conduction band for STO, as demonstrated in Fig. 4c. Electrons, coming from the occupied surface donor states, are then transferred to the empty conduction band of STO at the interface, creating the 2DEG. Until all the surface states are empty, in the ideal case, the Fermi level remains at the donor energy while more and more electrons are transferred with increasing the film thickness. This simple electrostatic model not only explains the formation of 2DEG in CZO/STO, but also yields the following expression of the critical barrier thickness, $t_c$:

$$t_c = [E_D - (E_{CBM}^{CZO} - E_{CBM}^{STO})]\varepsilon_0\varepsilon_{CZO} \Big/ eP_{CZO} \qquad (1)$$

where, $E_{CBM}^{CZO}$ and $E_{CBM}^{STO}$ are the conduction band minima (CBM) of CZO and STO, respectively. For $t>t_c$, the 2DEG density as a function of the CZO film thickness can be given by[22]:

$$en_s = P_{CZO}/(1 - t_c/t) \qquad (2)$$

Figure 4a summarized the CZO/STO sheet carrier density $n_s$ as a function of $t$. Note that the abrupt enhancement in $n_s$ at $t>15$ uc is probably due to the formation of large content of oxygen vacancies on the STO side due to interfacial redox reactions. This concern is supported by the facts that there is measureable conduction at the back side of the STO substrates when $t>15$ uc as well as that we observed an additional decay in sheet conduction resulting from oxygen absorption at the initial process for the post annealing in pure O$_2$ at 150 ºC of a $t=22$ uc sample (Supporting information, Fig. S3b). For this reason, we fitted the Equation (2) only to $t<15$ uc. A least-squares fit was achieved for $P_{CZO}=2.3\times10^{13}$ e/cm$^2$ and $t_c=6.5$ uc. The deduced value of $P_{CZO}=2.3\times10^{13}$ e/cm$^2$=3.5 $\mu$Ccm$^{-2}$, is in good agreement with the experimentally determined values for a comparable compressively strained 40 uc SrZrO$_3$/SrTiO$_3$ bilayer-heterostructure.[27] In short, the strain induced polarization model can fit well the critical thickness dependence of the sheet carrier density for CZO/STO heteostructures, and can explain largely the metallic conduction at the nonminally-nonpolar perovskite interface between the two band insulators, when the content of oxygen vacancies is negligible. It is noteworthy that the deduced polarization of $P_{CZO}=3.5$ $\mu$Ccm$^{-2}$ in CZO/STO ($t\leq14$ uc) is much lower than that expected at the polar LAO/STO interface, $P_{LAO}=e/2a_{LAO}^2=55.8$ $\mu$Ccm$^{-2}$. It, therefore, might be tantalizing to determine why the two different perovskite-type interfaces exhibit similar sheet carrier densities.

In conclusion, we have discovered a 2DEG at the sharp perovskite-type interface of CZO/STO with very high electron mobility exceeding 60,000 cm$^2$V$^{-1}$s$^{-1}$ at 2 K. The sheet carrier density of the 2DEG exhibits a critical thickness dependence, suggesting the polarization induced electronic reconstruction dominates the interface conduction. The strain-induced polarization can provide new avenues to explore high mobility 2DEGs at complex oxide interfaces.

## METHODS

**Sample growth and Characterizations.** The CaZrO$_3$ (CZO) films were grown on (001) TiO$_2$-terminated SrTiO$_3$ (STO) substrates (5×5×0.5 mm$^3$ with miscut less than 0.2º) by pulsed laser deposition (PLD) in an oxygen atmosphere of ~10$^{-4}$ mbar with the film growth process monitored by *in-situ* RHEED. During ablation, a KrF laser ($\lambda$=248 nm) with a repetition rate of 1 Hz and laser fluence of 1.5 Jcm$^{-2}$ was used. The target-substrate distance was fixed at 5.6 cm. The growth temperature was fixed at 600 ºC. A CZO ceramic pellet produced by Spark plasma sintering (SPS) was used as the target. Layer-by-layer two-dimensional growth of CZO films was optimized by RHEED oscillations with a growth rate of approximately 75 pulses per unit cell (0.05 Å/s) (Supplementary Information, Fig. S1a). After deposition, the sample was cooled under the deposition pressure with a rate of 15 ºC/min to room temperature (below 30 ºC). All samples are extremely stable at room temperature and can survive the annealing in 1 bar pure $O_2$ at 100 ºC for over 5 hours.

Electrical characterization was made mainly using a 4-probe Van der Pauw method with ultrasonically wire-bonded aluminum wires as electrodes. For all samples, the Hall resistance is linear with respect to magnetic field (0-2 T). The high resolution XRD measurements were performed on a Rigaku Smartlab system.

**STEM and EELS analysis.** Aberration-corrected STEM measurements were performed by an FEI Titan 80–300ST TEM equipped with a high brightness Shottky emitter (XFEG) and a Gatan Image Filter (Tridiem). High-angle annular dark field (HAADF) images were acquired at 300 kV, where the probe size, convergence angle and HAADF inner collection angle were 0.8–1.0 Å, 17.3 mrad and 71.9 mrad, respectively. For EELS in the STEM, an accelerating voltage of 120 kV (probe size of 1.5–2.0 Å) and the spectrum imaging technique were used to avoid changes in the specimen induced by the electron beam during acquistion. Spectrum images consisting of 40 ten-analysis-point lines parallel to the interface were acquired and an increment of 0.28 nm to next line was used. Each spectrum with an energy resolution of 0.9 eV was obtained at a dispersion of 0.1 eV/pixel for 0.2-0.4 s. Then the spectra along the lines were summed to increase signal-to-noise ratio.

**X-ray photoelectron spectroscopy (XPS) measurement**. The XPS measurements were performed in a Kratos Axis UltraDLD instrument, using a monochromatic Al Ka X-ray source with photon energy of 1,486.6 eV. This leads to a kinetic energy of Ti 2p electrons of roughly 1,025 eV. According to the NIST database(NIST Standard Reference Database 71,version 1.2), the electron escape depth is approximately 22Å in STO at this kinetic energy. The pass energy used for the high resolution scan was 20 eV. The detection angle of the electrons varied between 0º and 60º with respect to the sample normal.

**Acknowledgement.** The authors gratefully acknowledge the discussions with G. A. Sawatzky, T. S. Jespersen, L. P. Yu, N. Balke and the technical assistance from J. Geyti, M. Søgaard, and K.V. Hansen. Funding from the Danish Agency for Science, Technology and Innovation, is acknowledged.

**Supporting Information Available**. This material is available free of charge via the Internet at http://pubs.acs.org.

**REFERENCES AND NOTE:**

(1) Ohtomo, A.; Hwang, H. Y. *Nature* **2004**, *427*, 423.

(2) Chen, Y. Z.; Bovet, N.; Trier, F.; Christensen, D. V.; Qu, F. M.; Andersen, N. H.; Kasama, T.; Zhang, W.; Giraud, R.; Dufouleur, J.; Jespersen, T. S.; Sun, J. R.; Smith, A.; Nygård, J.; Lu, L.; Büchner, B.; Shen, B. G.; Linderoth, S.; Pryds, N. *Nat. Commun.* **2013**, *4:1371*, doi: 10.1038/ncomms2394.

(3) Mannhart, J.; Schlom, D. G. *Science* **2010**, *327*, 1607.

(4) Miletto Granozio, F.; Koster, G.; Rijnders, G. *MRS Bullettin* **2013**, *38*, 1017.

(5) Sulpizio, J. A.; Ilani, S.; Irvin, P.; Levy, J. *Annu Rev Mater Sci* **2014**, *44*, 117.

(6) Reyren, N.; Thiel, S.; Caviglia, A. D.; Fitting Kourkoutis, L.; Hammerl, G.; Richter, C.; Schneider, C. W.; Kopp, T.; Ruetschi, A.-S.; Jaccard, D.; Gabay, M.; Muller, D. A.; Triscone, J.-M.; Mannhart, J. *Science* **2007**, *317*, 1196.

(7) Richter, C.; Boschker, H.; Dietsche, W.; Fillis-Tsirakis, E.; Jany, R.; Loder, F.; Kourkoutis, L. F.; Muller, D. A.; Kirtley, J. R.; Schneider, C. W.; Mannhart, J. *Nature* **2013**, *502*, 528.

(8) Brinkman, A.; Huijben, M.; Van Zalk, M.; Huijben, J.; Zeitler, U.; Maan, J. C.; Van der Wiel, W. G.; Rijnders, G.; Blank, D. H. A.; Hilgenkamp, H. *Nat. Mater.* **2007**, *6*, 493.

(9) Lee, J. –S.; Xie, Y. W.; Sato, H. K.; Bell, C.; Hikita, Y.; Hwang, H. Y.; Kao, C. –C. *Nature Mater.* **2013**, *12*, 703.

(10) P. Irvin; Neazey, J. P.; Cheng G. L.; Lu, S. C.; Bark, C. W.; Ryu, S.; Eom, C. B.; Levy, J. *Nano Lett.*, **2013**, *13*, 364.

(11) Chen, Y. Z.; Zhao, J. L.; Sun, J. R.; Pryds, N.; Shen, B. G. *Appl. Phys. Lett.* **2010**, *97*, 123102.

(12) Santander-Syro, A. F.; Copie, O.; Kondo, T.; Fortuna,F.; Pailhes, S.; Weht, R.; Qiu, X. G.; Bertran, F.; Nicolaou, A.; Taleb-Ibrahimi, A.; Le Fevre, P.; Herranz, G.; Bibes, M.; Reyren, N.; Apertet,Y.; Lecoeur, P.; Barthelemy, A.; Rozenberg, M. J. *Nature* **2011**,469,189.

(13) Meevasana, W.; King, P. D. C.; He, R. H.; Mo, S-K.; Hashimoto, M.; Tamai, A.; Songsiriritthigul, P.; Baumberger, F.; Shen, Z-X. *Nat. Mater.* **2011**, *10*, 114.

(14) Plumb, N. C.; Salluzzo, M.; Razzoli, E.; Månsson, M.; Falub, M.; Krempasky, J.; Matt, C. E.; Chang, J.; Schulte, M.; Braun, J.; Ebert, H.; Minár, J.; Delley, B.; Zhou, K.-J.; Schmitt, T.; Shi, M.; Mesot, J.; Patthey, L.; Radović M. *Phys. Rev. Lett.* **2014**, *113*, 086801.

(15) Chen, Y. Z.; Pryds, N.; Kleibeuker, J. E.; Sun, J. R.; Stamate, E.; Koster, G.; Shen, B. G.; Rijnders, G.; Linderoth, S. *Nano. Lett.* **2011**, *11*, 3774.

(16) Kozuka, Y.; Kim, M.; Ohta, H.; Hikita, Y.; Bell, C.; Hwang, H. Y. *Appl. Phys. Lett.* **2010**, *97*, 222115.

(17) Hesper, R.; Tjeng, L. H.; Heeres, A.; Sawatzky, G. A. *Phys. Rev. B* **2000**, *62*, 16046.

(18) Chambers, S. A.; Engelhard, M. H.; Shutthanandan, V.; Zhu, Z.; Droubay, T. C.; Qiao, L.; Sushko, P. V.; Feng, T.; Lee, H. D.; Garfunkel, E.; Shah, A. B.; Zuo, J. –M.; Ramasse, Q. M. *Surf. Sci. Rep.* **2010**, *65*, 317.

(19) Kalabukhov, A.; Gunnarsson, R.; Börjesson, J.; Olsson, E.; Claeson, T.; Winkler, D. *Phys. Rev. B* **2007**, *75*, 121404.

(20) Yu L. P.; Zunger A. *Nat. Commun.* **2014**, *5:5118* doi: 10.1038/ncomms6118.

(21) Kleibeuker, J. E.; Zhong, Z.; Nishikawa, H.; Gabel, J.; Müller, A.; Pfaff, F.; Sing, M.; Held, K.; Claessen, R.; Koster, G.; Rijnders, G. *Phys. Rev. Lett.* **2014** *113*, 237402.

(22) Ibbetson J. P.; Fini, P. T.; Ness, K. D.; DenBaars, S. P.; Speck, J. S.; Mishra, U. K. *Appl. Phys. Lett.* **2000**, *77*, 250.

(23) Tampo, H.; Shibata, H.; Maejima, K.; Yamada, A.; Matsubara, K.; Fons, P.; Kashiwaya, S.; Niki, S.; Chiba, Y.; Wakamatsu, T.; Kanie, H. *Appl. Phys. Lett.* **2008**, *93*, 202104.

(24) Tsukazaki, A.; Ohtomo, A.; Kita, T.; Ohno, Y.; Ohno, H.; Kawasaki, M. *Science* **2007**, *315*, 1388.

(25) Bark, C. W.; Felker, D. A.; Wang, Y.; Zhang, Y.; Jang, H. W.; Folkman, C. M.; Park, J. W.; Baek, S. H.; Zhou, H.; Fong, D. D.; Pan, X. Q.; Tsymbal, E. Y.; Rzchowski, M. S.; Eom, C. B. *Proc. Natl. Acad. Sci. USA* **2011**, *108*, 4720.

(26) Kim, S.-I.; Kim, D.-H.; Kim, Y.; Moon, S. Y.; Kang, M.-G.; Choi, J. K.; Jang, H. W.; Kim, S. K.; Choi, J.-W.; Yoon, S.-J.; Chang, H. J.; Kang, C.-Y.; Lee, S.; Hong, S.-H.; Kim, J.-S.; Baek, S.-H. *Adv. Mater.* **2013** *25*, 4612.

(27) Tsurumi, T.; Harigai, T.; Tanaka, D.; Nam, S-M.; Kakemoto, H.; Wada, S.; Saito, K. *Appl. Phys. Lett.* **2004**, *85*, 5016.

(28) Yang, K.; Wang, C. L.; Li, J. C.; Zhao, M. L.; Wang, X. Y. *Solid State Commun.* **2006**, *139* 144.

(29) Huijben, M.; Koster, G.; Kruize, M. K.; Wenderich, S.; Verbeeck, J.; Bals, S.; Slooten, E.; Shi, B.; Molegraaf, H. J. A.; Kleibeuker, J. E.; van Aert, S.; Goedkoop, J. B.; Brinkman, A.; Blank, D. H. A.; Golden, M. S.; van Tendeloo, G.; Hilgenkamp, H.; Rijnders, G. *Adv. Funct. Mater.* **2013**, *23*, 5240.

(30) Xie, Y. W.; Bell, C.; Hikita, Y.; Harashima, S.; Hwang, H. Y. *Adv. Mater.* **2013,** *25*, 4735.

(31) Mathews, M. D; Mirza, E. B.; Momin, A. C. *J. Mater. Sci. Lett*. **1991**, *10*, 305.

(32) Cantoni, C.; Gazquez, J.; Miletto Granozio, F.; Oxley, M. P.; Varela, M.; Lupini, A. R.; Pennycook, S. J.; Aruta, C.; di Uccio, U. S.; Perna, P.; Maccariello, D. *Adv. Mater.* **2012**, 24, 3952.

(33) Nakagawa, N.; Hwang, H. Y.; Muller D. A. *Nat. Mater.* **2006**, *5*, 204.

(34) Herranz, G.; Basletic, M.; Bibes, M.; Carretero, C.; Tafra, E.; Jacquet, E.; Bouzehouane, K.; Deranlot, C.; Hamzic, A.; Broto, J.-M.; Barthelemy, A.; Fert, A. *Phys. Rev. Lett.* **2007**, *98*, 216803.

(35) Thiel, S.; Hammerl, G.; Schmehl, A.; Schneider, C. W.; Mannhart, J. *Science* **2006**, *313* 1942.

(36) Chen, Y. Z.; Bovet, N.; Kasama, T.; Gao, W. W.; Yazdi, S.; Ma, C.; Pryds, N.; Linderoth, S. *Adv. Mater.* **2014**, *26*, 1462.

(37) Tufte, O. N.; Chapman, P. W. *Phys. Rev.* **1967**, *155*, 796.

(38) Chen, Y. Z.; Pryds, N.; Sun, J. R.; Shen, B. G.; Linderoth, S. *Chin. Phys. B* **2013**, *22*, 116803.

(39) Sing, M.; Berner, G.; Goβ, K.; Müller, A.; Ruff, A.; Wetscherek, A.; Thiel, S.; Mannhart, J.; Pauli, S. A.; Schneider, C. W.; Willmott, P. R.; Gorgoi, M.; Schäfers, F.; Claessen, R. *Phys. Rev. Lett.* **2009**, *102*,176805.

**FIGURE CAPTIONS**.

**Figure 1. Atomically-flat epitaxially grown perovskite-type interface of CZO/STO. a**, Sketch of the CZO/STO heterostructure. **b**, AFM image (5 $\mu$m×5 $\mu$m) of a 50 uc CZO/STO heterostructure showing an atomically smooth surface. **c** and **d**, The XRD-RSM measurements in the vicinity of the (002) and (-103) perovskite Bragg peaks for the CZO/STO heterostructure.

**Figure 2. Sharp interface of the compressively strained CZO/STO heterostructure. a**, A HAADF-STEM image across the CZO/STO heterointerface; **b,** The averaged line profiles across the interface (A=Sr, Ca; B=Ti, Zr)**; c,** Relative lattice distortion of cations in the CZO film in the vicinity of interface; **d**, Sketch for the strain induced polarization due to the lattice distortion near the interface. The dot lines are guide to the eye.

**Figure 3. Highly mobile conduction at CZO/STO heterointerfaces. a-c,** The temperature dependence of sheet resistance, $R_s$, carrier density, $n_s$, and mobility, $\mu$, respectively. Most of the samples exhibit a large $\mu$ exceeding 20000 cm$^2$V$^{-1}$s$^{-1}$ at 2 K, despite the dramatic difference in $n_s$.

**Figure 4. Sketch of the formation of oxide 2DEG by strain-induced polarization**. **a**, Thickness dependent sheet carrier density; **b** and **c**, Band alignment before and after the formation of 2DEG, respectively, at the CZO/STO heterointerface.

**Table of Contents Graphic**

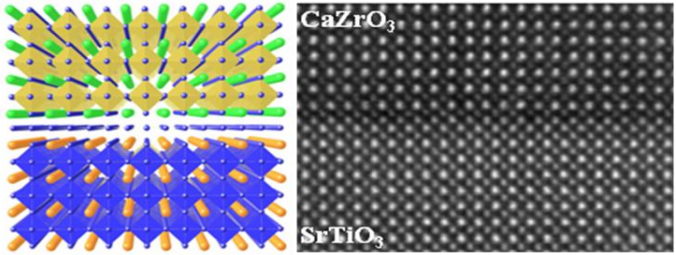

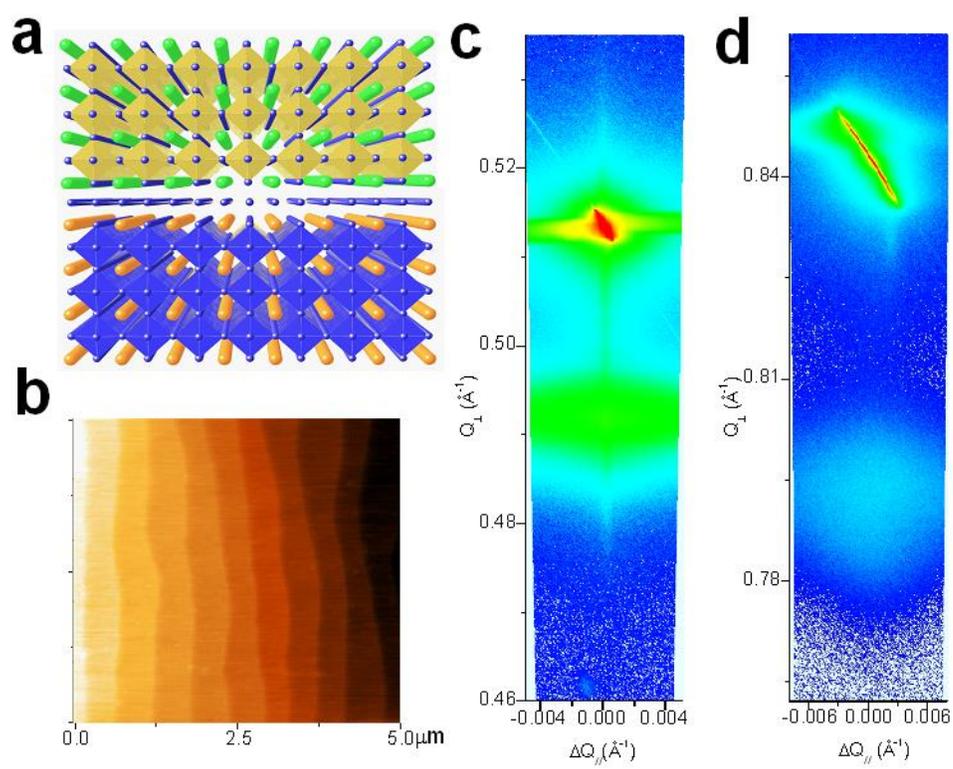

Fig. 1 Chen *et al.*

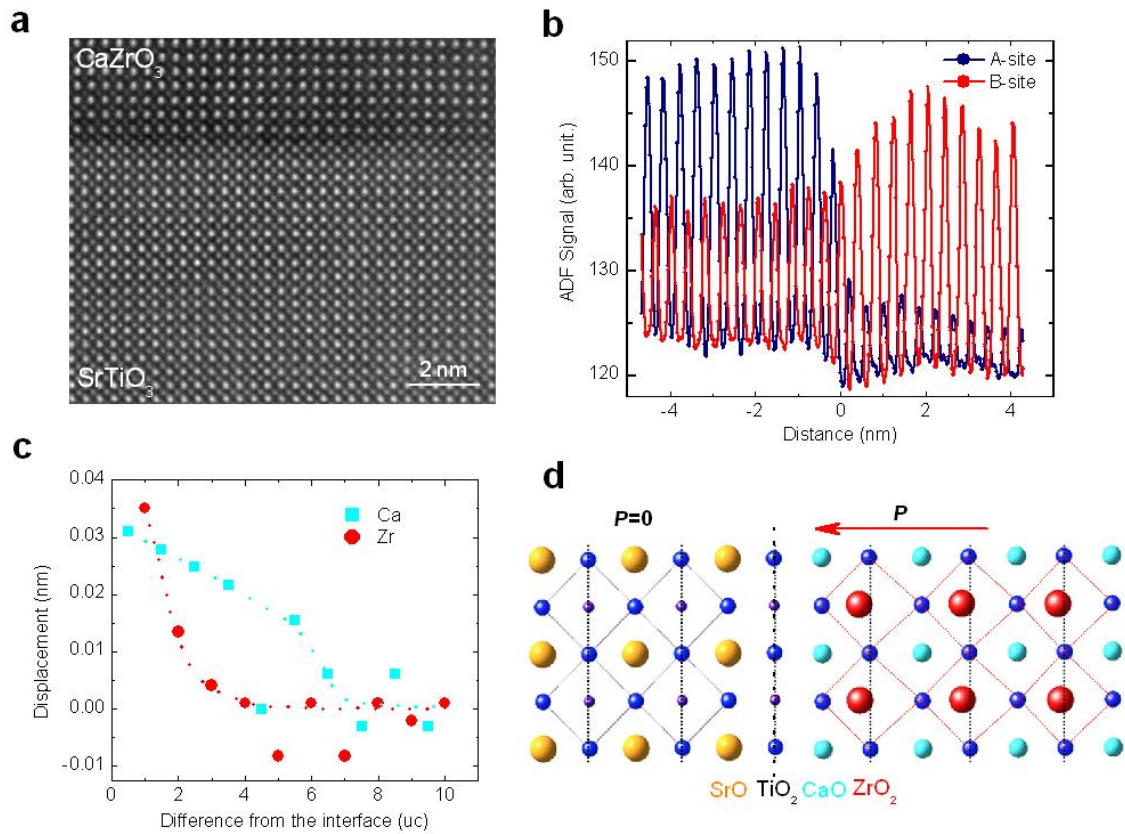

Fig.2 Chen *et al.*

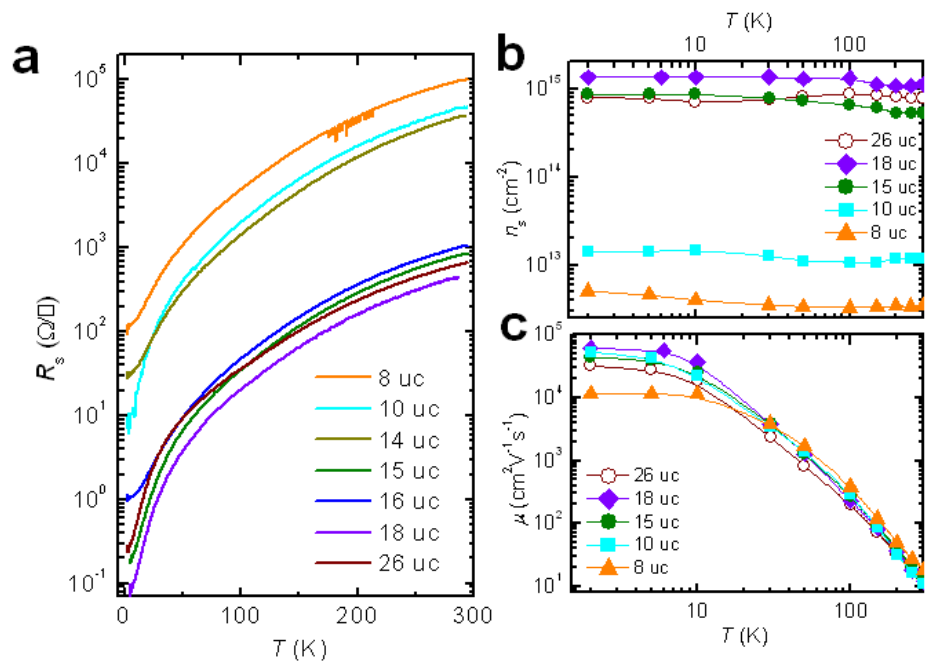

Fig. 3 Chen *et al.*

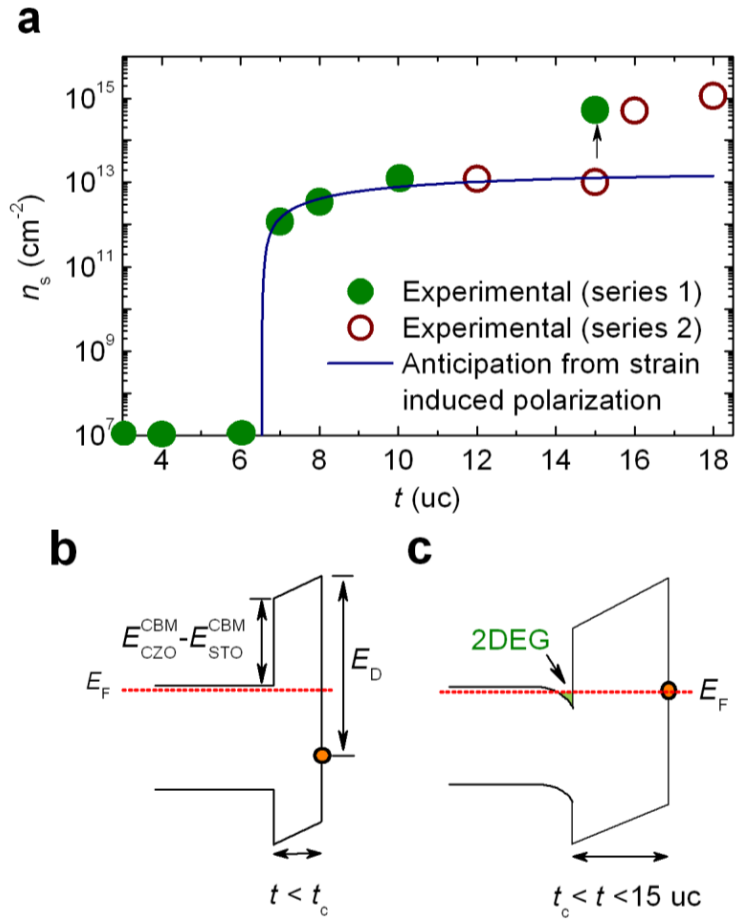

Fig. 4 Chen *et al.*